# Systematic literature review on forecasting and prediction of technical debt evolution

Ajibode Adekunle[a], Apedo Yvon[b], Ajibode Temitope[c]

[a]*School of Computing, Queen's University, Kingston, K7L 3N6, Ontario, Canada*
[b]*School of Computer Science, Northwestern Polytechnical University, Xi'an, 710072, Shaanxi, China*
[c]*Department of Computer Science, Federal College of Education (Technical), Akoka, 100213, Lagos, Nigeria*

## Abstract

**Context**: Technical debt (TD) refers to the additional costs incurred due to compromises in software quality, providing short-term advantages during development but potentially compromising long-term quality. Accurate TD forecasting and prediction are vital for informed software maintenance and proactive management. However, this research area lacks comprehensive documentation on the available forecasting techniques.
**Objective**: This study aims to explore existing knowledge in software engineering to gain insights into approaches proposed in research and industry for forecasting TD evolution.
**Methods**: To achieve this objective, we conducted a Systematic Literature Review encompassing 646 distinct papers published until 2023. Following established methodology in software engineering, we identified and included 14 primary studies for analysis.
**Result**: Our analysis unveiled various approaches for TD evolution forecasting. Notably, random forest and temporal convolutional networks demonstrated superior performance compared to other methods based on the result from the primary studies. However, these approaches only address two of the fifteen identified TD types, specifically Code debt and Architecture debt, while disregarding the remaining types.
**Conclusion**: Our findings indicate that research on TD evolution forecasting is still in its early stages, leaving numerous challenges unaddressed. Therefore, we propose several research directions that require further investigation to bridge the existing gaps.

∗For corresponding enquiries, contact ajibode.a@queensu.ca
*Email addresses:* ajibode.a@queensu.ca (Ajibode Adekunle), yvonapedo@mail.nwpu.edu.cn (Apedo Yvon), ajibodex@gmail.com (Ajibode Temitope)

## 1. Introduction

During software development and maintenance, it is not always possible for developers to produce optimal quality code that satisfies the specifications. Ward Cunningham (Cunningham, 1992) coined the term “Technical Debt” (TD) in 1992 to describe this imperfection. As in finance, where borrowing money is required to meet financial needs, developers must meet the requirements of a system under development. TD is a metaphor reflecting technical compromises that can yield short-term benefits, but may harm the long-term health of a software system (Li et al., 2015).

TD is not limited to the software implementation process, but is linked to the entire software development cycle (Brown et al., 2010) and can cause ambivalence towards software development. When incurred to achieve short-term benefits, TD can be productive (Allman, 2012) if its cost is visible and contained. Forecasting and predicting the TD during the evolution of a software application is an open and challenging research issue because both the software system and its TD emerge in parallel (Digkas et al., 2017). In most cases, TD forecasting are used interchangeably, which is the process of estimating the probability and potential amount of TD that could accumulate during the software development process. Both terms involve identifying and analyzing the factors that contribute to the emergence of TD and using that information to make informed decisions about software development (Mathioudaki et al., 2021).

However, some researchers have used the terms ’TD prediction’ and ’TD forecasting’ in slightly different ways. For instance, researchers like (Rantala and Mäntylä, 2020; Aversano et al., 2021)(Skourletopoulos et al., 2014) use the term ’TD prediction’ specifically for the application of machine learning or other quantitative techniques to predict the likelihood or severity of TD in software projects. On the other hand, some use the term ’TD forecasting’ more broadly to encompass any technique or method for estimating TD in software projects, including both quantitative and qualitative methods (Belle, 2019; Ampatzoglou et al., 2015; Seaman and Guo, 2011; Behutiye et al., 2017)(Mathioudaki et al., 2021). However, this study uses ’forecasting’ to represent both terms.

TD forecasting are important in the software development process, because they help software developers and project managers anticipate and plan for potential TD issues before they become major problems. By identifying factors that contribute to TD and estimating their potential impact, TD forecasting can help teams make informed decisions about software design, development, and maintenance.

Various tools can be used for forecasting TD, including software metrics, code analysis, machine learning, and expert judgment. Each technique has its own strengths and limitations, and the choice of technique depends on the specific context and goal of the software project. For example, software metrics can be used to quantify the complexity and quality of software code, whereas machine learning can be used to identify patterns and trends in software development data. Expert judgment relies on domain experts’ knowledge and experience to assess the potential impact of TD on software quality and performance.

Despite the challenges and complexities involved in TD forecasting, it is an important research area with the potential to significantly improve the quality and sustainability of software applications. Numerous reviews have been conducted on various aspects of TD management (Li et al., 2015; Lenarduzzi et al., 2021; Rios et al., 2018), self-admitted TD (Sierra et al., 2019), TD Prioritization (Alfayez et al., 2020), TD identification (Alves et al., 2016), and TD estimation tools (Tsoukalas et al., 2018). However, to the best of our knowledge, the existing literature relating to TD lacks a study that summarizes the current literature on TD evolution forecasting that this study considers.

This leads us to formulate various research questions ($RQ_s$) aimed at comprehensively understanding the existing body of literature concerning the forecasting of TD evolution, as explained below:

**$RQ_1$:** What are the current approaches for forecasting TD?

Our aim is to understand the state-of-the-art techniques and methodologies employed in the field of TD evolution forecasting. By examining the existing literature, we aim to identify the various models and algorithms that researchers have used to forecast the evolution of TD in software applications.

**$RQ_2$:** Which types of TD do the forecasting and predicting approaches address?

We investigate the types of TD addressed by TD forecasting approaches, such as code debt, architectural debt, and requirement debt. Through an examination of existing research, we can determine the types of TD that have been the primary focus of forecasting efforts. This analysis provides insight into the specific challenges and considerations associated with each type.

**$RQ_3$:** What are the common projects, project artifacts, and programming languages used in the forecasting of TD?

We explore the common projects, artifacts, and programming languages utilized in forecasting and predicting TD. By analyzing the literature, we can identify the types of projects that have been the primary focus of forecasting studies and the programming languages predominantly used. This information provides insights into the applicability and generalizability of existing research findings.

**$RQ_4$:** Which metrics are predominantly used in TD forecasting studies?

We investigate the metrics predominantly employed in TD forecasting studies. Understanding which metrics are most frequently used provides insight into the key factors considered by researchers when predicting TD evolution.

**$RQ_5$:** Which evaluation metrics are predominantly used for measuring the performance of TD forecasting models?

We investigate the evaluation metrics commonly used to measure the performance of TD forecasting models, such as precision, recall, F1 score, and area under the receiver operating characteristic curve (AUC-ROC). Identifying the evaluation metrics predominantly utilized in the field sheds light on current practices and standards for assessing the reliability and effectiveness of TD forecasting models.

Reviewing existing TD evolution forecasting studies is crucial to gain a better understanding of how software systems evolve over time and how TD can affect the evolution process. Such a review can help researchers and practitioners to identify gaps in the existing literature and suggest future research directions in this field.

Therefore, the objective of this study is to conduct a systematic literature review (SLR) (Kitchenham, 2012) to analyze the current state of TD evolution forecasting approaches. An SLR involves collecting data from previously published studies on a specific topic, analyzing and interpreting the data, summarizing the findings, and drawing a detailed conclusion (Kitchenham et al., 2015). An SLR will enable us to systematically identify, collect, and analyze relevant literature, including primary studies and secondary sources, to provide a thorough understanding of the research area. By synthesizing and summarizing the findings of multiple studies, we can identify the strengths and limitations of different approaches, highlight key research trends, and identify research gaps that require further investigation. The objectives of this SLR study on TD forecasting are as follows.

- To understand the current state of research on TD forecasting.
- To identify the type of TD that researchers are predicting and forecasting.
- To determine which metrics are commonly studied to predict and forecast TD.
- To identify the techniques that are frequently used for forecasting and predicting TD.
- To identify the source of datasets that are frequently used for forecasting and predicting TD.
- To identify promising directions for future research on TD forecasting.

The remainder of this paper is structured as follows: Section 2 provides background information, Section 3 outlines the methodology used for the SLR, Section 4 and 5 analyze and discuss the obtained results, Section 6 discusses the potential threats to the validity of this research, and finally, Section 7 presents the conclusions and suggests areas for future studies.

## 2. Background

This section introduces the concepts explored in this research, along with relevant existing works.

### *2.1. Technical Debt*

The notion of TD was initially introduced by Cunningham in 1992, defining it as "the debt incurred by speeding up software project development, resulting in deficiencies that lead to high maintenance overheads" (Cunningham, 1992). Seaman et al. (2011) further elaborated on TD, describing it as incomplete or immature artifacts within the software development lifecycle, which contributes to increased costs and diminished quality (Seaman and Guo, 2011). Although these artifacts may expedite short-term development, their low quality often leads to the long-term expenses associated with maintenance efforts and corrections.

McConnell (2008) refined the definition of TD as "a design or construction approach that offers short-term expediency but creates a technical context where future work will cost more than it would if done presently, including increasing costs over time" (McConnell, 2008). Avgeriou et al. (2016) expanded on this definition by stating that TD encompasses design or implementation constructs that provide short-term advantages but establish a technical context that can make future changes more costly or even impossible. They emphasized that TD represents a liability, both actual and potential, primarily affecting internal system qualities, such as maintainability and evolvability (Avgeriou et al., 2015). However, Ernst (2021) asserted that TD is more than just badly written code. It refers ultimately to a mismatch between what the software should have been and what it actually is (Ernst et al., 2021).

Additionally, Li et al. (2015) conducted a systematic mapping study to gain a comprehensive understanding of TD and generated an overview of the current research on its management. Based on their analysis of 96 selected studies, they proposed a classification of ten types of TD at different levels (Li et al., 2015).

Similarly, McConnell (2008) categorized TD into two types: unintentional TD, which occurs involuntarily and nonstrategically, often due to poorly planned activities by inexperienced professionals or changes in the environment, and intentional TD, which is deliberate and strategically motivated, involving decisions by professionals to achieve short-term benefits through shortcuts, alternative solutions, or deferred tasks (McConnell, 2008). However, it is worth noting that most prediction studies in this area are based on unintentional TD.

In a study by Rios et al. (Rios et al., 2018), TD manifested in various activities and phases of the software development life cycle. The authors identified 15 distinct types of TDs, which are detailed in Table I along with their respective definitions.

Table I: Types and definition of TD

| TD Type | Definition |
|---|---|
| Code Debt | Denotes issues discovered in the source code, such as poorly written code that violates best coding practices or rules, which can hinder code readability and maintenance. |
| Test Debt | Refers to problems identified during testing activities that can affect the quality of those activities. |
| Documentation Debt | Represents issues found in software project documentation. |
| Infrastructure Debt | Pertains to infrastructure issues that, if present in the software organization, can impede or delay development activities, adversely affecting the team's ability to deliver a high-quality product. |
| Design Debt | Describes debt that becomes evident when analyzing the source code and identifying violations of good object-oriented design principles. |
| Requirements Debt | Indicates the trade-offs made regarding which requirements the development team should implement or how to implement them. In essence, it reflects the gap between the optimal requirements specification and the actual system implementation. |
| People Debt | Refers to people-related issues that, if present in the software organization, can hinder or delay development activities. |
| Build Debt | Represents issues that make the build task more challenging and unnecessarily time-consuming. |
| Defect Debt | Signifies known defects, typically identified through testing activities or user reports in bug tracking systems, which the development team acknowledges should be fixed but are deferred due to competing priorities and limited resources. |
| Process Debt | Describes inefficient processes, where the existing process may no longer be suitable for its intended purpose. |
| Automation Test Debt | Denotes the effort required to automate tests for previously developed functionality to support continuous integration and faster development cycles. |
| Usability Debt | Signifies inappropriate usability decisions that will require adjustments later on. |
| Service Debt | Refers to the incorrect selection and substitution of web services, leading to a mismatch between the service features and the requirements of the applications. This type of debt is particularly relevant for systems with service-oriented architectures. |
| Versioning Debt | Refers to issues related to source code versioning, such as unnecessary code forks. |
| Architecture Debt | Encompasses problems encountered in product architecture that can impact architectural requirements. Typically, architectural debt arises from suboptimal initial solutions or solutions that become suboptimal as technologies and patterns become outdated, compromising internal quality aspects like maintainability. |

Since this identification stems from a recent secondary study and represents the most comprehensive compilation available in the literature, we adopted these types of TD for our study.

Therefore, taking into account the various perspectives mentioned above regarding TD, in this study, we define TD as *the repercussions of shortcuts, hasty decisions, and suboptimal practices during software development, resulting in accumulated consequences, leading to a higher cost of maintenance, decreased productivity, and potential difficulties in future development and scalability.*

### *2.2. forecasting Concept*

Forecasting involves the art of predicting future occurrences based on an analysis of historical and current data, typically by observing patterns. The ability to anticipate the future values of a given characteristic is important across various scientific and engineering disciplines (Palit and Popovic, 2006). Over the years, owing to the growing diversity and complexity of forecasting problems, numerous techniques have been devised and continue to be developed, each of which is tailored to serve a specific purpose. The field of forecasting has long been shaped by statistical methods, which can be broadly categorized into two types: causal (or associative) and time series models. Causal models, including the commonly used regression analysis, operate on the assumption that a cause-and-effect relationship exists between the variable of interest and other variables. Consequently, these models aim to uncover such relationships in order to forecast future values. On the other hand, time-series models, such as the widely employed ARIMA model, treat the analyzed system as a mysterious entity and posit that the necessary information for forecasting is embedded within a sequence of time-dependent data, which is expected to follow the same patterns observed in the past (Das, 2012).

However, in recent decades, the forecasting community has witnessed the emergence of Machine Learning (ML) models, which have garnered considerable attention and positioned themselves as formidable alternatives to classical statistical models (Bontempi et al., 2013). These ML models, also referred to as black boxes or data-driven models, employ self-correcting learning algorithms that leverage supervised, unsupervised, or reinforcement learning techniques to acquire knowledge regarding the stochastic relationship between past and future events, relying solely on historical data. Experts' opinions regarding the superiority of either time-series or ML approaches in terms of prediction accuracy differ. In a recent study conducted by Makridakis et al. (Makridakis et al., 2018), the authors contended that ML methods must enhance their accuracy, reduce computational requirements, and shed their black-box nature. Their paper makes a significant contribution by demonstrating that traditional statistical methods outperform ML methods and emphasizes the need to investigate the underlying reasons while seeking ways to rectify the situation. However, it should be noted that their comparisons acknowledge the possibility that the results were influenced by the specific dataset used. They argued that if the time series are considerably longer, ML methods have the potential to optimize their weightings more effectively. Conversely, Christy demonstrated that Artificial Neural Networks (ANNs) can yield superior outcomes compared to traditional statistical methods such as linear regression and Box-Jenkins (ARMA, ARIMA) approaches (Christy et al., 2022). More recently, other models have emerged, including regression trees, support vector regression, and nearest neighbor regression (Friedman et al., 2001; Alpaydin, 2010).

### *2.3. Existing SLR works*

In this Section, we briefly report on previous systematic reviews (Systematic Mapping Studies and Systematic Literature Reviews) available in the literature and show their main goals in Table II. We present these studies in chronological order to show the research evolution of TD.

Table II: Previous SLR study

| Year | Paper Title | Citation | Goal |
|---|---|---|---|
| 2021 | A systematic literature review on Technical Debt prioritization: Strategies, processes, factors, and tools | (Lenarduzzi et al., 2021) | Understanding technical debt prioritization |
| 2021 | Characterizing Technical Debt and Antipatterns in AI-Based Systems: A Systematic Mapping Study | (Bogner et al., 2021) | TD characterization and antipatterns |
| 2020 | A systematic literature review of technical debt prioritization | (Alfayez et al., 2020) | Tools for prioritizing and Managing TD |
| 2019 | A survey of self-admitted technical debt | (Sierra et al., 2019) | Causes and consequences of self-admitted technical debt |
| 2019 | Automated measurement of technical debt: A systematic literature review | (Khomyakov et al., 2019) | Tools for quantifying and assessing TD |
| 2018 | A tertiary study on technical debt: Types, management strategies, research trends, and base information for practitioners | (Rios et al., 2018) | Types, management strategies, research trends in TD |
| 2018 | Managing architectural technical debt: A unified model and systematic literature review | (Besker et al., 2018) | Architectural TD management |
| 2017 | Analyzing the concept of technical debt in the context of agile software development: A systematic literature review | (Behutiye et al., 2017) | TD in agile software development |
| 2017 | Identification and analysis of the elements required to manage technical debt by means of a systematic mapping study | (Fernández-Sánchez et al., 2017) | TD element requirements |
| 2016 | Identification and management of technical debt: A systematic mapping study | (Alves et al., 2016) | TD identification and management |
| 2016 | Decision criteria for the payment of technical debt in software projects: A systematic mapping study | (Ribeiro et al., 2016) | Decision criteria for TD payment |
| 2015 | The financial aspect of managing technical debt: A systematic literature review | (Ampatzoglou et al., 2015) | Costs and benefits associated with TD |
| 2015 | A systematic mapping study on technical debt and its management | (Li et al., 2015) | TD management and classification |
| 2013 | An exploration of technical debt | (Tom et al., 2013) | Nature and implications of TD |

Previously, Lenarduzzi et al. (Lenarduzzi et al., 2021) examined existing research and industry practices in software engineering to explore different approaches proposed for prioritizing TD. After reviewing 557 literature sources, they focused on 44 primary studies. The findings revealed a variety of TD prioritization approaches, each with distinct objectives and optimization criteria. However, the study highlighted a lack of validated tools for TD prioritization and identified a preliminary state of research with no consensus on the important factors and measurement methods. The authors emphasized the need for further investigations to address these gaps and provide guidance for future TD prioritization studies.

Bogner et al. (Bogner et al., 2021) conducted an examination and analysis of various types of TD that manifest in artificial intelligence (AI)-based systems. They also explored the antipatterns and corresponding proposed solutions associated with these TD types. The researchers identified four novel forms of TD, namely data debt, model debt, configuration debt, and ethics debt. Additionally, they discovered a total of 72 antipatterns linked to deficiencies in data and model aspects within AI-based systems.

Alfayez et al. (Alfayez et al., 2020) examined the interdependencies of software artifacts and the extent of required human involvement. They assessed prioritization approaches based on their considerations of value, cost, or resource limitations.

In their study, Khomyakov et al. (Khomyakov et al., 2019) explored available tools for measuring and analyzing TD, focusing on quantitative methods that could be automated. Out of the 331 papers retrieved, they carefully selected 21. Their findings have revealed the emergence of numerous novel approaches for TD measurements.

Rios et al. (Rios et al., 2018) conducted a comprehensive investigation encompassing five research questions and evaluated 13 secondary studies spanning from 2012 to March 2018. They developed a taxonomy of TD categories, identified situations in which debt items manifested in software projects, and created a visual representation illustrating the current state of activities, strategies, and tools for supporting TD management. Their findings shed light on areas within TD research that warrant further investigation, including the identification of management activities that lack appropriate supporting tools.

Besker et al. (Besker et al., 2018) delved into the realm of Architectural Technical Debt (ATD), amalgamating and synthesizing research endeavors to generate new insights specifically focused on ATD. They examined publications from 2005 to 2016, ultimately selecting 43 relevant studies. The results underscored the absence of comprehensive guidelines for successfully managing ATD in practical settings as well as the absence of an integrated process encompassing these activities.

Behutiye et al. (Behutiye et al., 2017) conducted a comprehensive analysis of the state-of-the-art of TD within the context of agile software development (ASD). Their study investigated the causes, consequences, and management strategies for TD. The researchers reviewed relevant publications until 2017 and carefully selected 38 studies for their analysis. Through this examination, they identified potential research areas that warrant further investigation. The study highlighted the significant interest in TD and its relationship with ASD while also providing insights into specific categories that often contribute to TD, such as a focus on rapid delivery and architectural and design issues.

Fernández-Sánchez et al. (Fernández-Sánchez et al., 2017) aimed to identify the key elements necessary for effective TD management. The researchers considered relevant publications until 2017 and meticulously selected 69 studies for their analysis. While the study did not provide an overarching overview of the TD phenomenon or associated management activities, it successfully classified the identified elements into three distinct groups based on stakeholders' perspectives: engineering, engineering management, and business-organizational management. The groups encompassed fundamental decision-making factors, cost estimation techniques, and decision-making practices and techniques.

Alves et al. (Alves et al., 2016) conducted an extensive investigation of strategies proposed for the identification and management of TD in software projects. Their study focused on publications between 2010 and 2014, and the researchers systematically selected 100 studies for analysis. This study proposed an initial taxonomy of TD types and presented a comprehensive list of indicators for identifying TD and corresponding management strategies.

Furthermore, the researchers analyzed the current state of TD research, shedding light on potential research gaps. These findings indicate a growing interest among researchers in the field of TD. Specifically, this study identified gaps in the proposal of new indicators, management strategies, and tools for controlling TD. Additionally, empirical studies are required to validate the proposed strategies.

Ribeiro et al. (Ribeiro et al., 2016) undertook an evaluation to determine the appropriate timing for addressing TD items and the application of decision-making criteria to strike a balance between short-term benefits and long-term costs. The researchers reviewed pertinent publications until 2016 and meticulously selected 38 studies for their analysis. The study identified 14 decision-making criteria that development teams can employ to prioritize the resolution of TD items. In addition, a comprehensive list of TD types associated with these criteria was provided.

Ampatzoglou et al. (Ampatzoglou et al., 2015) conducted a systematic literature review to examine research efforts related to TD in the context of software engineering, with a particular emphasis on financial aspects. They screened 69 studies published until 2015 and found that a clear mapping between financial and software engineering concepts is still lacking. As a contribution to this field, they developed a glossary of terms and a classification scheme for financial approaches that can be applied to managing TD.

Li et al. (Li et al., 2015) performed a systematic mapping study on TD management (TDM) to provide a comprehensive view of the current state of research on TDM. They screened 94 studies published between 1992 and 2013 and presented a classification of TD concepts. This study identified several gaps in the literature related to the TDM process, such as the need for high-quality empirical studies, the application of TDM approaches in industrial contexts, and the development of effective tools for managing different types of TD during the TDM process.

Tom et al. (Tom et al., 2013) conducted an exploratory case study using a multivocal literature review and interviews with software practitioners and academics to define the scope of TD. Through this process, they developed a comprehensive theoretical framework that encompasses the various dimensions, attributes, precedents, and outcomes of TD. This framework offers a valuable perspective for practitioners seeking to comprehend the broader TD phenomenon and its practical implications.

In contrast to previous Systematic Literature Reviews (SLRs), our study is the first to systematically investigate existing research studies that discuss the forecasting of TD evolution.

Our study provides a comprehensive and critical overview of the current state of research on the forecasting of TD. By synthesizing and analyzing the findings of existing studies, our research identifies the strengths and limitations of current approaches, highlights research gaps, and suggests potential avenues for future research. This can aid in developing a more nuanced and evidence-based understanding of the factors that influence the evolution of TD over time, and can ultimately help software teams to effectively manage and mitigate TD in their projects.

## 3. Study design

We conducted a systematic literature review to gain insights into current research on TD evolution forecasting. Our approach followed the guidelines established by Kitchenham and Charters (Kitchenham and Brereton, 2013) as well as Kitchenham and Charters (Kitchenham and Charters, 2007). In addition, we utilized the “snowballing” process defined by Wohlin (Wohlin et al., 2012).

The initial phase entails meticulous planning, encompassing the formulation of research questions, development of a robust search strategy, determination of search sources, formulation of precise search strings, identification of selection criteria, and definition of quality assessment parameters.

Subsequently, a thorough selection process ensued, whereby we conducted comprehensive searches across various databases to identified unique and pertinent studies. Relevant studies were identified and carefully selected based on predefined criteria. To ensure a comprehensive approach, the inclusion of all relevant studies is confirmed utilizing an advanced AI tool, connected paper[1], mitigating the possibility of inadvertently omitting crucial literature. Additionally, the quality of the selected studies was rigorously evaluated by considering factors such as methodological robustness, credibility of data sources, and overall research design.

Upon completing the selection process, the information required to address the research questions was extracted from the selected studies. This involves a meticulous examination of the full text, enabling the systematic extraction of pertinent data.

Subsequently, the extracted data were subjected to rigorous analysis. Statistical and qualitative analysis techniques were employed to derive meaningful insights and identify patterns or trends within the dataset. Following the analysis, a comprehensive discussion ensues wherein the findings are critically evaluated in light of the existing literature. Interpretations and explanations of observed patterns or outcomes are provided, ensuring a well-rounded exploration of the research topic.

Finally, recommendations are formulated based on the analysis and discussion. These recommendations serve to guide future research and inform relevant stakeholders of potential implications or courses of action. Furthermore, areas for future research were identified, highlighting avenues for further investigation.

By adhering to this systematic and scientific framework, our review ensures a robust and comprehensive examination of the research topic, provides valuable insights, and contributes to the advancement of scientific knowledge in the field.

Similarly, we provide an overview of the goals and research questions, along with a detailed description of our search strategy. Furthermore, we evaluated the quality of each included paper and outlined the data extraction and analysis process.

### *3.1. Search strategy*

The search strategy involved identifying the most pertinent bibliographic sources and search terms, establishing inclusion and exclusion criteria, and implementing a selection process to guide inclusion decisions. Within our search string, we incorporated all terms associated with TD, as proposed by Rios et al. (Rios et al., 2018) and documented in Table I.

**Search Terms:** The search string contained the following terms:

We used the asterisk (*) as a wildcard character to capture various word endings and variations, aiming to enhance the chances of discovering publications that address the forecasting of TD evolution. This search strategy was applied to the titles, abstracts, and keyword fields in the selected databases.

[1] https://www.connectedpapers.com/

Table III: Search Terms

| Search Term | Keywords |
|---|---|
| Technical Debt | technical debt AND (predict* OR foreca*) |
| Design Debt | design debt AND (predict* OR foreca*) |
| Code Debt | code debt AND (predict* OR foreca*) |
| Architecture Debt | architecture debt AND (predict* OR foreca*) |
| Test Debt | test debt AND (predict* OR foreca*) |
| Defect Debt | defect debt AND (predict* OR foreca*) |
| Documentation Debt | documentation debt AND (predict* OR foreca*) |
| Infrastructure Debt | infrastructure debt AND (predict* OR foreca*) |
| Requirement Debt | requirement debt AND (predict* OR foreca*) |
| People Debt | people debt AND (predict* OR foreca*) |
| Build Debt | build debt AND (predict* OR foreca*) |
| Process Debt | process debt AND (predict* OR foreca*) |
| Automation Debt | automation debt AND (predict* OR foreca*) |
| Usability Debt | usability debt AND (predict* OR foreca*) |
| Service Debt | service debt AND (predict* OR foreca*) |
| Versioning Debt | versioning debt AND (predict* OR foreca*) |

**Bibliography sources:** Regarding the selection of bibliographic sources, we adhered to the recommendations of Kitchenham and Charters (Kitchenham and Charters, 2007), as these sources are widely recognized as representative within the software engineering domain and are commonly employed in reviews. The selected sources encompassed the IEEE Xplore Digital Library, Science Direct, ACM Digital Library, Scopus, Wiley Online Library, Google Scholar, and Springer Link. Furthermore, we conducted a manual search of prominent conferences and workshops related to TD, such as the International Workshop on Managing Technical Debt (MTD) and the International Conference on Technical Debt (TechDebt), due to their specific focus on TD.

We established inclusion and exclusion criteria for the title, abstract, and full text, as outlined in Table IV.

Table IV: Inclusion and exclusion criteria

| **Criteria** | **Evaluation criteria** | **Applied to** |
|---|---|---|
| Inclusion | Paper propose a model or technique for forecasting or predicting TD evolution in software system | Title, abstract, and full text |
| | Paper used some measures to forecast or predict when TD should be repayed to avoid TD mismanagement | Title, abstract, and full text |
| | Paper identified the source of dataset used | Title, abstract, and full text |
| | Paper is empirically validated | Title, abstract, and full text |
| Exclusion | Papers not written in English | Title and Abstract |
| | Papers not peer-reviewed | Title and Abstract |
| | Duplicated papers (we removed the older version and consider the latest version) | Title and Abstract |
| | Study that the library can not locate its full text | Title and Abstract |
| | Study only mentions TD in an introductory statement and does not focus on its prediction or forecasting | Full text |
| | Study that mentions forecast or predictions in an introductory statement and does not focus on software TD | Full text |
| | Study is an editorial, keynote, opinion, tutorial, workshop summary report, poster, or panel. Such papers have been excluded either due to their small size or due to the fact that such articles are usually not peer-reviewed | Title and abstract |
| | Studies that are out of the scope of this research, such as TD identification, Management of TD, Papers that discuss financial and market debt, all papers not related to software TD but have the word debt or forecast or predictions | Full text |
| | Papers that discusses financial debt | Full text |

**Search and selection process:** The literature search was first conducted in May 2023 and then reconducted in October 2023. The latter was to discover if there were any other latest primary studies in the field, encompassing all available publications up to those points. For Google Scholar, we executed the entire search string, resulting in the extraction of 294 unique papers. In the case of the IEEE Xplore library, due to limitations in searching all keywords simultaneously, we divided the search string into two sets. We first conducted a search using the first set of strings, downloaded the corresponding papers, and then repeated the process with the second set. This approach involved partitioning the items in Table III into two equal parts to accommodate the library's constraints, resulting in a total of 78 unique papers.

In the ACM Digital Library, we conducted a search using the complete string, leading to the download of 92 unique papers. In the SpringerLink database, we used the complete string while employing the filter command to restrict the search results to articles and conference papers, yielding 65 unique papers. On the Wiley Online Library, we utilized all the search strings and limited the search to journals, aligning with our desire to focus on scholarly, peer-reviewed articles. Journals often publish rigorous research studies, academic reviews, and contributions from experts in the field. By restricting our search to journals, we aimed to ensure a high standard of research quality and relevance to the study, compared to other types of publications like books and reference works. We downloaded 74 unique papers in this library.

In the case of the Scopus database, a similar approach to that used for the IEEE Xplore library was adopted due to difficulties encountered when searching for a complete string. Consequently, 43 unique papers were downloaded for further examination.

In summary, a total of 646 papers were initially selected for further reading, comprising findings from various databases and sources as outlined above.

**Test-running the applicability of inclusion and exclusion:** To evaluate the suitability of the inclusion and exclusion criteria, a preliminary assessment was performed on a randomly selected subset of fifteen papers (assigned to all authors) from the retrieved papers as suggested by Kitchenham and Brereton (Kitchenham and Brereton, 2013).

**Applying exclusion criteria to title and abstract:** The refined criteria were used to evaluate the 646 papers. Four authors reviewed each paper, and in cases of disagreement, the fifth author was consulted to resolve any conflicts. However, no disagreement arose between the authors during the evaluation. Out of the initial 646 papers, 66 were included based on their title and abstract.

**Full reading:** Thoroughly examining the 66 papers that met the inclusion criteria based on their title and abstract, we employed the criteria outlined in Table IV and assigned each paper to four authors for a comprehensive review. As a result of this process, we identified 22 papers that exhibited potential relevance as valuable contributions.

**Snowballing:** Employing the snowballing technique by Wohlin (Wohlin, 2014), we thoroughly explored the references cited within the retrieved papers and examined all papers that referenced those retrieved papers. This meticulous process has yielded an additional relevant paper. We followed the same approach used for the retrieved papers during the snowball search, which was conducted in October 2023. In total, we identified 15 additional papers with potential relevance, but ultimately only one of them met the criteria to be included in the final set of publications.

**Connected Paper:** In order to ensure thoroughness in our research, we utilized the remarkable capabilities of “connected paper.” Connected Paper is a cutting-edge AI-based website that serves as an indispensable resource for researchers. It provides a comprehensive platform for exploring and analyzing relevant academic papers in a highly efficient and systematic manner. By harnessing the power of artificial intelligence, connected papers assist researchers in navigating the vast landscape of scholarly literature, ensuring that no crucial papers are overlooked. This invaluable tool allowed us to meticulously examine a substantial number of relevant papers, numbered between 35 and 41. We meticulously reviewed the titles, keywords, and abstracts of these papers in a sequential manner. Importantly, we ensured that our study accounted for all pertinent connected papers for every original paper title searched, leaving no stones unturned in our pursuit of relevant literature.

As documented in Table V, our search and selection process yielded a total of 23 papers that were retrieved for the review.

Table V: The outcomes of the search process, selection of papers, and the rigorous application of quality assessment criteria.

| Steps | Total papers | Rejected papers |
|---|---|---|
| Retrieved publications (unique papers) | 646 | - |
| First reading (title and abstract) | 66 | 580 |
| Full reading | 22 | 44 |
| Backward and forward snowballing | 1 | 0 |
| Connected paper tool | 0 | 0 |
| Paper Identified | 23 | 0 |
| Quality assessment | 14 | 9 |
| **Primary studies** | **14** | |

### *3.2. Quality Assessment*

Prior to commencing the review, the quality of the selected papers was assessed to determine their suitability for supporting our research objective. Following the protocol proposed by Dybå and Dingsøyr (Dybå and Dingsøyr, 2008), we used a checklist Table VI consisting of specific questions to evaluate the chosen papers. Each question was assigned a ranking on a three-point Likert scale (No = 0, Partially = 0.5, Yes = 1), following the methodology employed by Said et al. (Said et al., 2020).

Table VI: Quality assessment (QA) checklist and scoring

| $QA_s$ | Item |
|---|---|
| $QA_1$ | Is the statement of the problem clearly defined? |
| $QA_2$ | Is the contribution of the study clearly defined? |
| $QA_3$ | Is the technique, model, or method clearly validated? |
| $QA_4$ | Are the limitation and future directions clearly stated? |
| $QA_5$ | Does the study primarily focus on the prediction or forecasting of TD? |
| Scores for the item | |
| No = 0, Partially = 0.5, Yes = 1 | |

The maximum achievable score was five, and only papers with a score of two or above were deemed acceptable. To ensure reliability, the evaluation process was independently repeated by another author. Ultimately, only papers with a cumulative score 4.5 or higher were selected. This meticulous approach aimed to exclusively include primary studies pertinent to the forecasting of TD.

From the initial search and selection process, a total of 23 papers were included in the review. However, upon applying the aforementioned quality assessment criteria, only 14 papers satisfied the required standards, as detailed in Table V.

### *3.3. Data Extraction*

After finalizing the selection of the primary studies, the full texts were obtained and downloaded. Subsequently, a thorough data extraction form was created and employed to collect the essential information. The details extracted using this form, along with the mapping of the information relevant to each research question, are summarized in Table VII.

Table VII: The extracted data for answering the $RQ_s$

| $RQ_s$ | Extracted data | Expected outcome |
|---|---|---|
| $RQ_1$ | Used model | linear regression, negative binomial regression, random forest, decision tree, etc |
| $RQ_2$ | Types of TD studied | Test, Documentation, Infrastructure, Design, Requirement, etc |
| $RQ_3$ | TD dataset | Commons-bcel, Commons-beanutils, Commons-cli, Commons-collections, Camel, Log4J, Hadoop, etc |
| $RQ_4$ | TD metrics list | Total number of variables, Total number of modifiers, Total number of returns, Total number of number used, Total number of loops, Max nested blocks, Seniority of Developers, etc. |
| $RQ_5$ | TD forecasting model evaluation metrics | Accuracy, Precision, Recall, etc |

Distinct approaches were employed to gather data pertaining to each research question (RQ). To address $RQ_1$, the names of the models used in each study were extracted to ascertain their respective approaches. The prevailing methods encompass a range of approaches including machine learning, statistical analysis, natural language processing, and deep learning. This investigation aimed to discern the prevalent approach in the literature. The implication of this research question is that by identifying the prevalent approach in the literature, we can gain insights into the dominant methodologies used to address TD evolution forecasting, thereby contributing to the advancement of scientific understanding in the field.

To address $RQ_2$, we systematically identify and analyze various types of TD evolution explored in prior research. These types include code debt and architectural debt. The main objective of this analysis was to gain a comprehensive understanding of how each proposed methodology in the existing research is applied to forecast and predict the evolution of each specific type of TD. Furthermore, this analysis aims to identify areas of research in which certain types of TD have not yet been adequately addressed.

To investigate $RQ_3$, we recorded the constituent projects comprising the datasets utilized in each study, along with their respective sources. In addition, we documented the programming languages employed to construct the datasets for each study. This information provides insights into the shared datasets employed to forecast and predict the evolution of TD.

The literature encompasses several TD metrics. However, not all of these metrics are applicable to predicting and forecasting the evolution of TD. Hence, $RQ_4$ aims to assess the array of TD metrics employed to forecast and predict the evolution of TD. This evaluation enables us to identify the most effective metrics for accurately predicting and forecasting TD evolution in software systems.

Empirically, when developing a forecasting and predicting model, it is customary to assess its performance and generalization. Thus, $RQ_5$ seeks to identify the evaluation metrics that have proven to be the most effective in studies related to forecasting and predicting the evolution of TD. The implications of this analysis are significant in terms of accurately gauging the reliability and effectiveness of forecasting models, ultimately aiding informed decision-making and proactive management of TD in software systems.

## 4. Study Results

In this section, we report the evidence found in the systematic literature review.

### *4.1. Demography Result*

In this study, we identified 66 primary studies, also referred to as primary research (PR). The earliest study on TD forecasting (Tsoukalas et al., 2019) was published in May 2019, while the most recent one (Ardimento et al., 2022) was published in December 2022. It is noteworthy that (Aversano et al., 2023) was actually published in November 2022 but is cited with the year 2023. After including and excluding PR, we selected 14 of these studies for analysis.

The findings of this study indicate a shift in TD evolution forecasting over the last five years: 2019 (Fontana et al., 2019; Tsoukalas et al., 2019), 2020 (Mhawish and Gupta, 2020; Rantala and Mäntylä, 2020; Tsoukalas et al., 2020), 2021 (Tsoukalas et al., 2021; Mathioudaki et al., 2021; Aversano et al., 2021), and 2022 (Mathioudaki et al., 2022; Garcia et al., 2021; Aversano et al., 2023; Zozas et al., 2022; Lerina et al., 2022; Ardimento et al., 2022).

Our analysis reveals that the majority of forecasting and prediction of TD evolution took a turn in 2022, with over 42% of the analyzed studies published in that year, compared to only 14% in 2019. This trend highlights the growing attention and interest in the research community towards understanding and predicting the evolution of technological disruption. The substantial increase in publications related to TD forecasting by 2022 underscores the growing recognition of the importance of this field. It is also important to note that there is no study on TD forecasting in the year 2023 at all. This suggests that the field may be currently underrepresented or that researchers have not yet explored TD forecasting within the context of the most recent year.

Furthermore, as shown in Figure I, the majority of the PRs analyzed in this study were published in recent years, specifically from 2019 to 2022, with over 40% published in 2022 (Figure Ic). $PR_s$ are accessible through multiple online resources, enabling further research and reading. Notably, $PR_{12}$ appeared in four different libraries, making it the most prevalent study among the PRs. However, it is important to note that the number of appearances does not necessarily correlate with higher citation counts of other researchers. Similarly, $PR_1$, which has the highest percentage of citations, is found in three online resources (Figure Ia). This percentage of citation is calculated as the number of $PR_1$ over all the citations of all the PRs in this study. This finding suggests that the number of online libraries in which a publication appears does not significantly determine its citation impact.

The $PR_s$ analyzed in this study are available on seven distinct online resources: IEEE Explorer, Springer Link, Science Direct, Google Scholar, Wiley Online, Scopus and ACM. However, the ScienceDirect, Google Scholar, and Springer Link databases host the largest proportion of $PR_s$ and demonstrate higher citation counts than other online resources (Figure Ib). Researchers often access $PR_s$ conveniently through Researchgate. It is important to note that the enhanced citation count cannot be attributed solely to the number of online resources in which a publication is accessible.

In addition, we examined whether the percentage of citations of freely accessible $PR_s$ exceeds that of closed-access $PR_s$. Interestingly, our study revealed a higher percentage of freely accessible $PR_s$, which correlates significantly with enhanced PR citations.

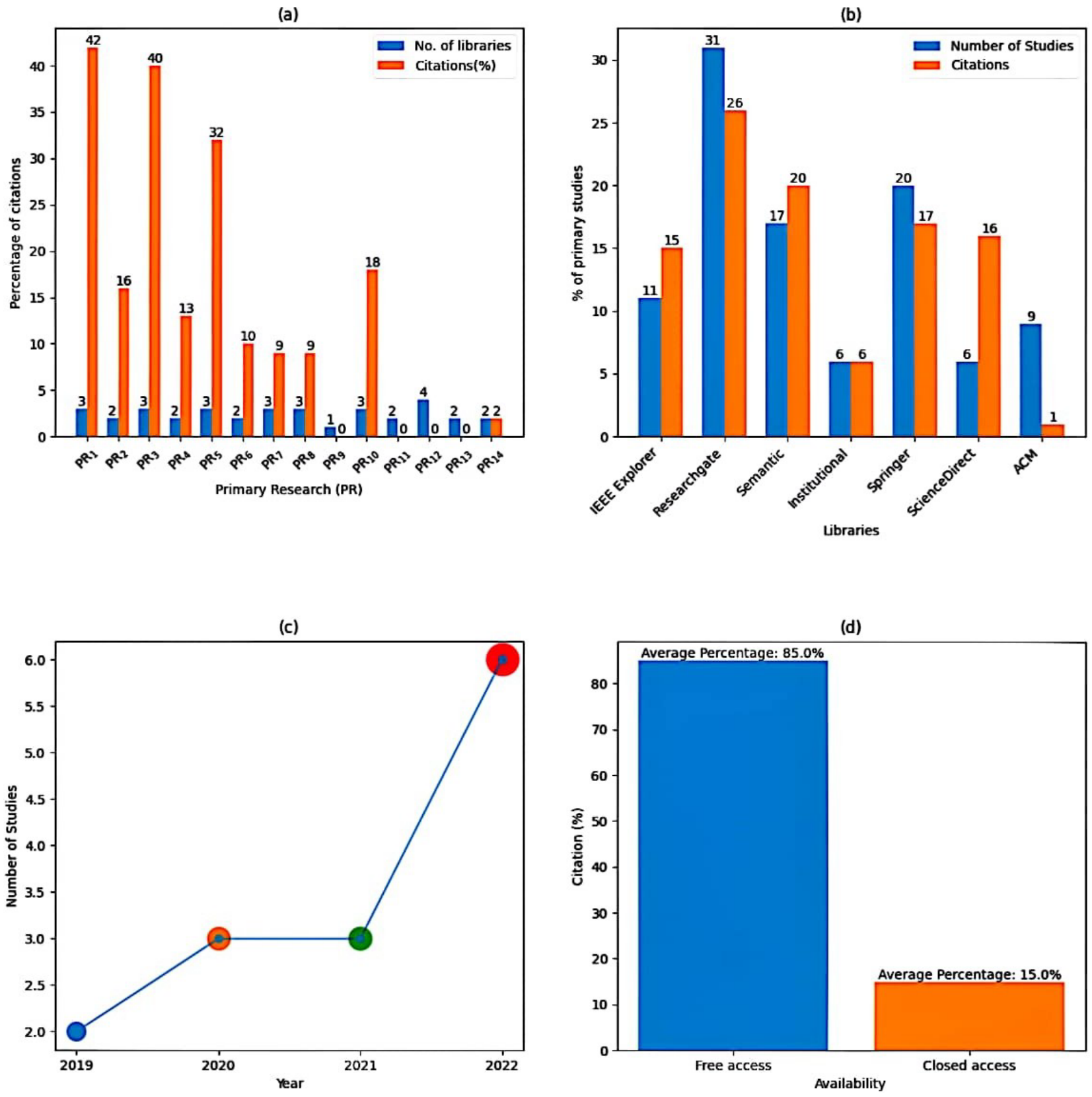

Figure I: (a) The number of libraries in which each PR appears and the cumulative citation; (b) The percentage of PR from each venue and cumulative citations; (c) The number of publications per year; (d) The percentage of free and closed access PR.

The contributions of researchers involved in the examined primary studies are represented in Figure II, constructed using Networkx, a robust Python library for network analysis. A total of 28 active researchers have made substantial contributions to this field. Noteworthy figures like Dimitrios Tsoukalas, Miltiadis Siavvas, and Dionysios Kehagias actively engage in predicting and forecasting the evolution of TD in software systems, demonstrating their significant contributions and collaborations with other researchers. This highlights a collaborative and interconnected research network within the relatively young field.

The implications of these results suggest the importance of considering factors other than the number of online resources when assessing the impact and citation counts of $PR_S$. Accessibility and collaboration among researchers significantly contribute to the visibility and recognition of their work. Moreover, the availability of freely accessible $PR_S$ appears to positively influence citation rates, highlighting the potential benefits of open-access publications in driving scientific impact.

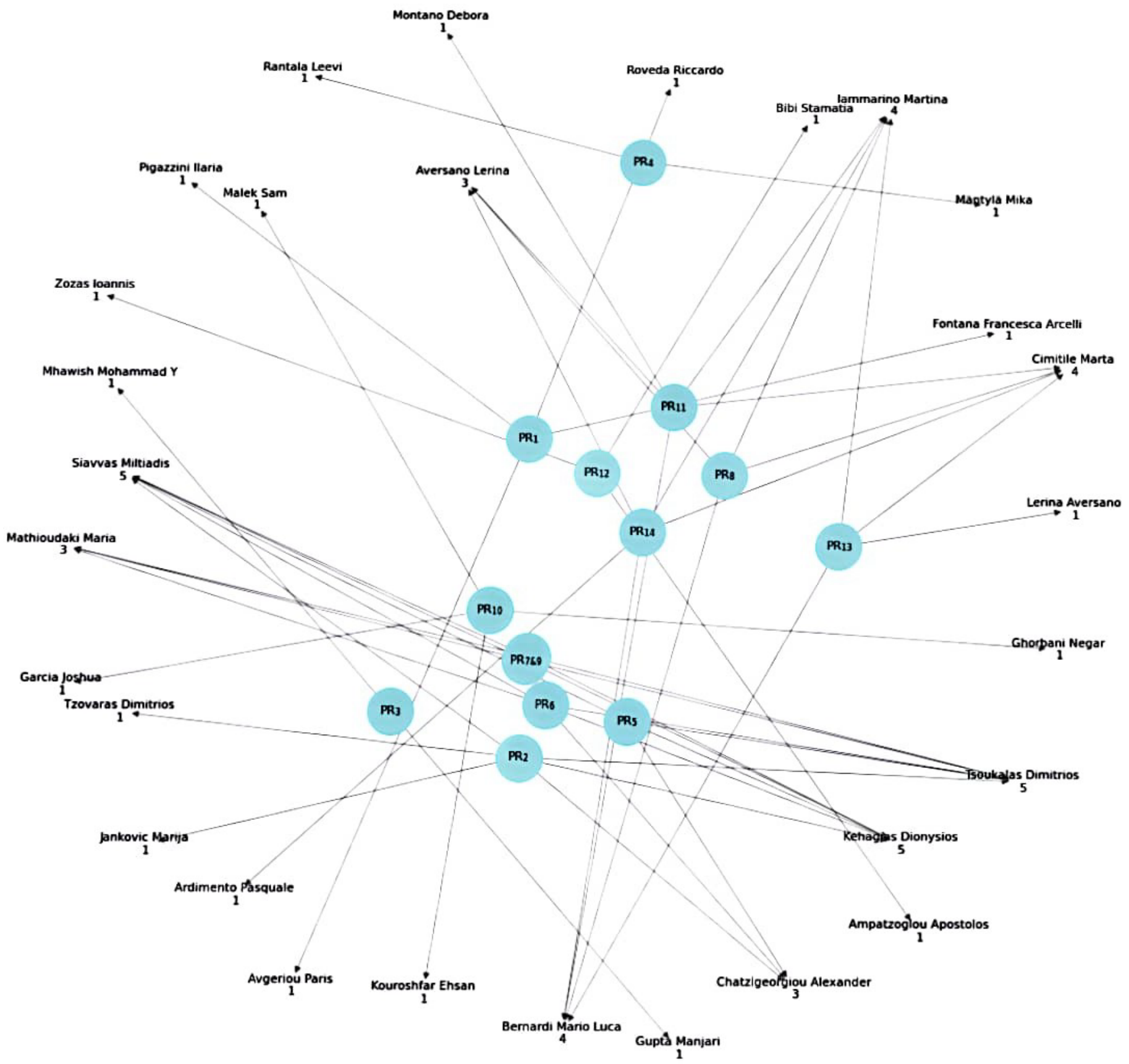

Figure II: Contributions of each researcher to the PRs

### *4.2. RQ$_1$: What are the current approaches for forecasting and predicting TD?*

In this research question, we analyze the various approaches employed to predict and forecast the evolution of software systems in terms of TD. Figure III(a) reveals that a total of 23 approaches have been utilized and are categorized into four different categories as depicted in Table VIII: Statistical Learning, Machine Learning, Time Series Analysis, and Natural Language Processing.

Among these approaches, more than 42% of existing studies have utilized RF for forecasting TD evolution, as shown in Table VIII. Notably, most studies employ multiple approaches rather than relying solely on a single one.

When considering the 23 employed models, Figure III(b) illustrates that 10 approaches—Lasso, SVM, ARIMAX, NBR, RF, JSTD tool, ARIMA (0,1,1), BoW, MPA, and TCN—demonstrate superior performance over others in the literature. This determination was made by analyzing how frequently these approaches outperformed others in the primary studies' papers, where papers employing multiple approaches explicitly mentioned which approach performed the best. Studies emphasizing the effectiveness of RF and TCN, belonging to the Machine Learning category, outnumber those that focus on other approaches, indicating the consistent effectiveness of Machine Learning models in addressing the evolution of TD.

It is important to highlight that some studies have employed only one approach, whereas others have utilized multiple approaches. In cases where a single approach was employed, it was assumed that the chosen approach performed best within that study. Conversely, if multiple approaches are employed, the one that demonstrates the best performance is identified and selected.

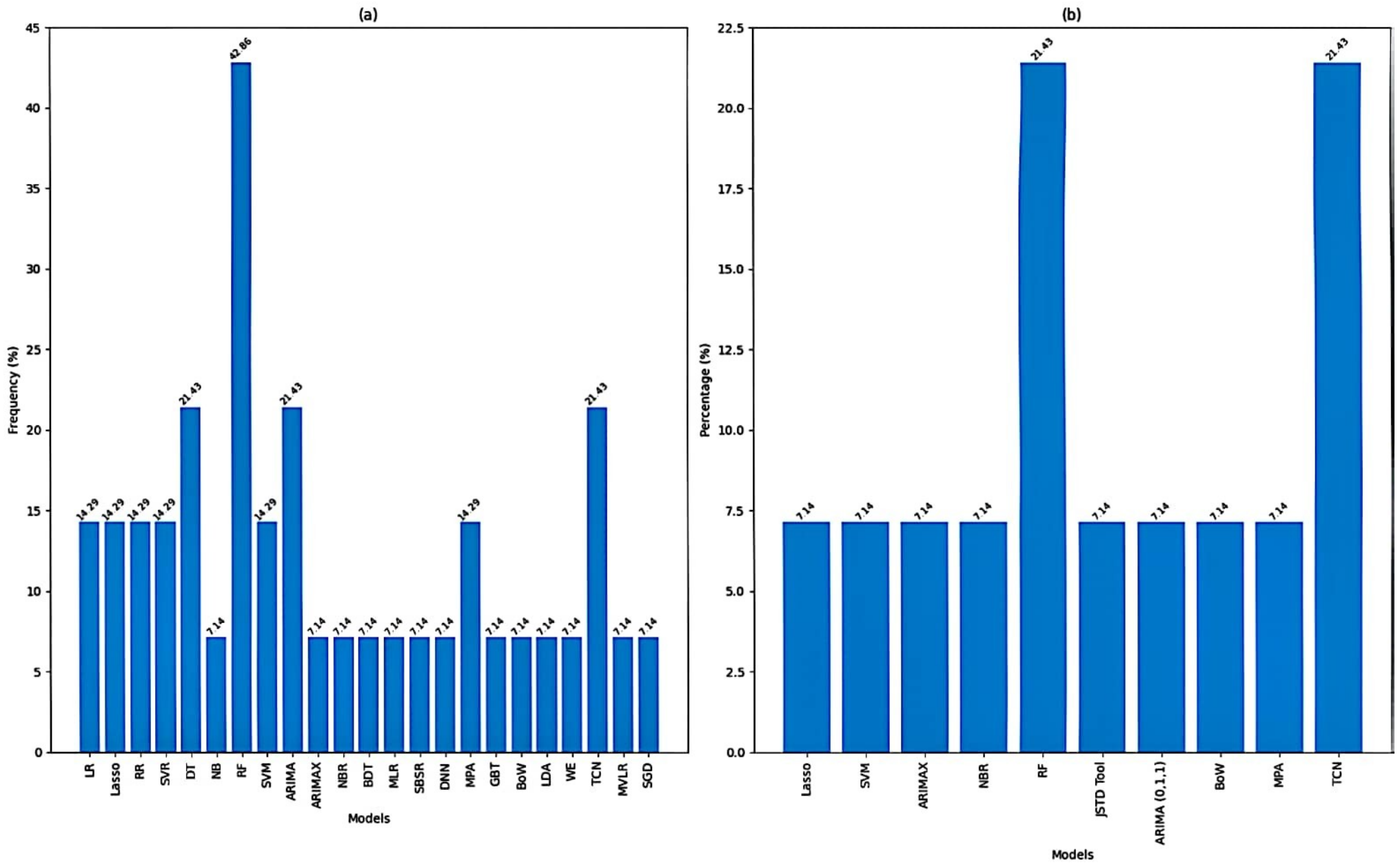

Figure III: Current approaches for predicting and forecasting TD

*Answer to $RQ_1$*
*Currently, 23 approaches are utilized for forecasting TD evolution, spanning Statistical Learning, Machine Learning, Time Series Analysis, and Natural Language Processing (NLP) methodologies. Within the Machine Learning category, the techniques from NLP methodologies are integral. More than 42% of existing studies have utilized Random Forests (RF) for forecasting TD evolution, as shown in Table VIII. Our analysis, considering the performance across multiple approaches in primary studies, identified 10 approaches—Lasso, SVM, ARIMAX, NBR, RF, JSTD tool, ARIMA (0,1,1), BoW, MPA, and TCN—as consistently demonstrating superior performance over others. Studies emphasizing the effectiveness of RF and TCN, belonging to the Machine Learning category, outnumber those that focus on other approaches, indicating their consistent effectiveness in addressing the evolution of TD.*

### *4.3. $RQ_2$: Which types of TD do the forecasting and predicting approaches address?*

It is crucial for researchers to propose methods that are capable of forecasting the evolution of each type of debt. This research question (RQ) aimed to evaluate the extent to which these debt types have been addressed by researchers with the intention of facilitating the work of software engineers.

According to Figure IV, out of the 15 identified debt types, only two have been addressed thus far: Code debt and Architecture debt. Among these, Code debt has received the most attention, with 87.5% of existing studies proposing methodologies to predict and forecast its evolution in software systems. In contrast, only 12.5% of the studies addressed architectural debt. However, the other 13 types of TD with 0% each have not received any attention thus far. The implication of these findings is that there is a significant gap in the research addressing the remaining 13 debt types identified by Rios et al. This lack of attention poses challenges for software engineers to effectively manage and mitigate the potential risks associated with these types of debt.

Table VIII: Categorization of Approaches

| Category | Approaches |
|---|---|
| Statistical Learning | Linear Regressor (LR), Lasso Regressor (Lasso), Ridge Regressor (RR), Negative Binomial Regression (NBR), Multiple Linear Regression (MLR), Supervised Backwards Stepwise Regression (SBSR), Multivariate Linear Regression (MVLR) |
| Machine Learning | Support Vector Regressor (SVR), Decision Tree (DT), Na¨ıve Bayes (NB), Random Forests (RF), Support Vector Machines (SVM), Bagged Decision Tree (BDT), Deep Neural Network (DNN), Multilayer Perceptron Algorithms (MPA), Gradient Boosted Trees (GBT), Temporal Convolutional Network (TCN), Stochastic Gradient Descent (SGD) |
| Time Series Analysis | ARIMA, ARIMAX |
| Natural Language Processing | Bag-of-Words (BoW), Latent Dirichlet Allocation (LDA), Word Embeddings (WE) |

Understanding and predicting the evolution of each debt type is crucial for maintaining software quality, minimizing TD accumulation, and ensuring the long-term sustainability of software systems. Neglecting the study of these debt types can lead to various negative consequences such as increased maintenance costs, decreased system performance, and reduced development productivity.

*Answer to $RQ_2$*
*Forecasting approaches have primarily addressed code debt and architecture debt, while the remaining 13 types of TD have not received any attention, indicating a significant research gap that poses challenges for software engineers in managing and mitigating associated risks, potentially leading to negative consequences for software quality and development productivity.*

### 4.4. *$RQ_3$: What are the common projects, project artifacts, and programming languages used in the forecasting and prediction of TD?*

To address this research question, we conducted a thorough analysis of existing studies to identify software application projects that are utilized as datasets for forecasting and predicting the evolution of TD. In addition, we investigated the sources of these projects, including the programming languages employed.

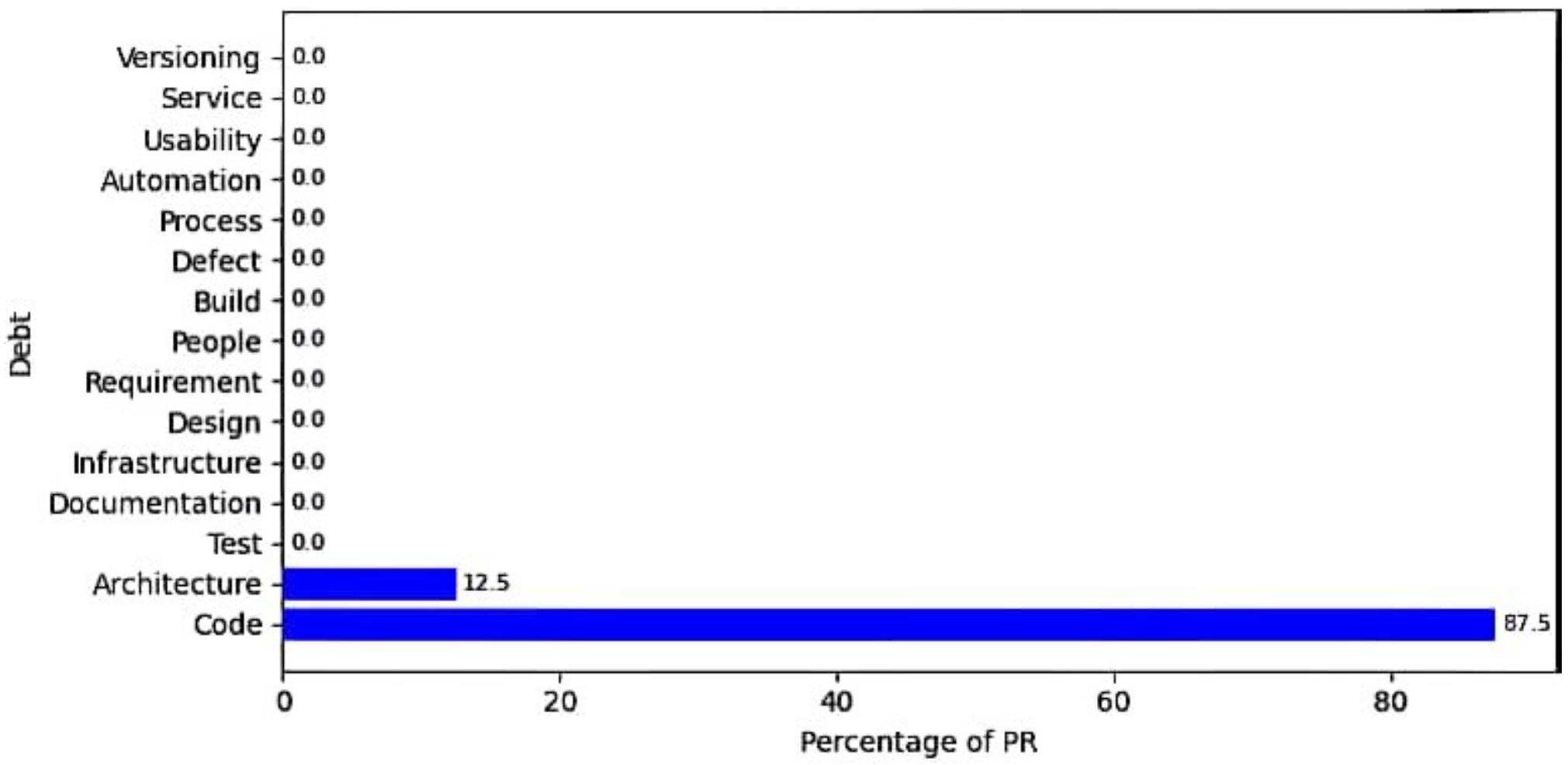

Figure IV: Types of TD addressed so far in the PRs

It is crucial to acknowledge that the dataset presented in the provided table does not encompass the benchmark utilized by Zozas et al. (Zozas et al., 2022), which was taken from Amanatidis et al. (Amanatidis et al., 2020), consisted of 105 benchmarks and was published on Zenodo[2]. Furthermore, we excluded the benchmarks employed by Mohammad et al. (Mhawish and Gupta, 2020), who utilized 111 benchmarks in their study and made the dataset available on their GitHub repository[3].

We present the findings in this study in Table IX, which shows the number of project benchmarks utilized for forecasting the evolution of TD in research, along with the corresponding number of studies that employed each benchmark. The data revealed that 66 software benchmarks were used to evaluate the performance of the TD evolution forecasting models. Among these benchmarks, only 10 have been employed in more than three studies, namely Commons-io, Commons-codec, Zookeeper, Jackson-dataformat-xml, Mina-sshd, Common-imaging, Jfreechart, Httpcomponents-client, Jackson-core, and Square OkHttp. It is important to note that all these projects were implemented using the Java programming language, except for the study conducted by Zozas et al. (Zozas et al., 2022), which has multiple languages. This suggests that the generalization and applicability of the proposed TD evolution forecasting models may be limited to software benchmarks written in specific programming languages, such as Java and JavaScript, as other programming languages have not been extensively studied in this context.

*Answer to $RQ_3$:*
*The common projects employed for TD evolution forecasting, including Commons-io, Commons-codec, Zookeeper, Jackson-dataformat-xml, Mina-sshd, Common-imaging, Jfreechart, Httpcomponents-client, Jackson-core, and Square OkHttp, are all written in the Java programming language.*

[2]https://zenodo.org/record/3979784
[3]https://github.com/zozas/jstd/blob/main/Manuscript%20dataset.zip

Table IX: Benchmarks used in predicting and forecasting TD evolution

| Various Datasets | | |
|---|---|---|
| Commons-io (6) | Beam (1) | Gerrit (1) |
| Commons-codec (4) | Camel (1) | Groovy (1) |
| Common-imaging (3) | Camel (Cam) (1) | Guava (1) |
| Httpcomponents-client (3) | Cassandra (Cas) (1) | Guice (1) |
| Jackson-core (3) | Commons-Math (1) | HBase (H) (1) |
| Jackson-dataformat-xml (3) | Commons-bcel (1) | Hadoop (1) |
| Jfreechart (3) | Commons-cli (1) | Hive (Hi) (1) |
| Mina-sshd (3) | Commons-collections (1) | Httpcomponents-core (1) |
| Square OkHttp (3) | Commons-configuration (1) | Incubator-dubbo (1) |
| Zookeeper (3) | Commons-daemon (1) | JGit (1) |
| Ambari (2) | Commons-dbcp (1) | JUnit4 (1) |
| Apache Kafka (2) | Commons-dbutils (1) | JavaWebSocket (1) |
| Apache SystemML (2) | Commons-digester (1) | Javassist (1) |
| Commons-beanutils (2) | Commons-exec (1) | Log4J (1) |
| Commons-net (2) | Commons-fileupload (1) | Nifi (1) |
| Jenkins (2) | Commons-jelly (1) | OpenJPA (Op) (1) |
| Square Retrofit (2) | Commons-jexl (1) | Santuario (1) |
| Xerces2-j (2) | Commons-jxpath (1) | Springboot (1) |
| Apache Tomcat (1) | Commons-ognl (1) | Tomcat (1) |
| Atlas (1) | Commons-validator (1) | Zxing (1) |
| Aurora (1) | Commons-vfs (1) | Apache/Ofbiz (1) |
| Batik (1) | Felix (1) | Igniterealtime/Openfire (1) |

### 4.5. *RQ$_4$: Which metrics are predominantly used in TD prediction and forecasting studies?*

The identification of metrics predominantly employed by researchers for forecasting TD evolution constitutes a critically significant endeavor that can greatly assist researchers in their subsequent experimental endeavors. To carefully collect and document the various metrics used in TD forecasting, we implemented a systematic approach. A dedicated spreadsheet was created to catalog each metric encountered in the selected primary studies. This spreadsheet served as a dynamic repository that was regularly updated as new metrics were identified during the review process. The meticulous tracking of metrics involved counting the instances of each metric's appearance in the reviewed papers. Within the scope of this study, we successfully ascertained 153 metrics that are associated with TD and that have been utilized in the forecasting of TD.

Table X: Existing metrics used in forecasting of TD evolution

| | | | |
|---|---|---|---|
| Complexity (9) | Commits for Every Release (2) | Halstead Effort, HPV Halstead Program Volume (1) | Number of Interfaces (1) |
| Line of Code (9) | Cyclic Dependency (2) | Halstead Program Level Difficulty (1) | Number of Local Variables (1) |
| Depth of Inheritance Tree (8) | Number of Attributes (2) | Hub-like Dependency (1) | Number of Message Chain Statements (1) |
| Bugs (7) | Number of Children (2) | Implicit Cross Package Dependency (1) | Number of Non-Accessors Methods (1) |
| Code Smells (7) | Number of Files Owners (2) | Incoming Module Dependency (1) | Number of Non-Constructor Methods (1) |
| Coupling Between Objects (7) | Number of Function Parameters (2) | Inheritance Coupling (1) | Number of Non-Final and Non-Abstract Methods (1) |
| Vulnerabilities (7) | Number of Functions (2) | Inner-module Co-changes (1) | Number of Non-Final and Static Attributes (1) |
| Lack of Cohesion in Methods (6) | Number of Public Methods (2) | Internal Module Dependencies (1) | Number of Non-Final and Static Methods (1) |
| Number of Methods (6) | Owned File Ratio (2) | Keyword Statement (1) | Number of Not Accessor or Mutator Methods (1) |
| Response for a Class (6) | Ownership of the Commit (2) | Line of Code Without Accessor or Mutation Method (1) | Number of Not Final and Non-Static Attributes (1) |
| Weight Method Count per Class (6) | Reliability Remediation Effort (2) | Link Overload (1) | Number of Obfuscation Incidents (1) |
| Comment Lines (5) | SQALE Rating (2) | Locality of Attribute Accesses (1) | Number of Open Issues (1) |
| Number of Classes (5) | Security Remediation Effort (2) | Maintainability Remediation Effort (1) | Number of Overridden (1) |
| Number of Static Invocations (4) | Seniority of Developers (2) | Maximum Message Chain Length (1) | Number of Packages (1) |
| SQALE Index (4) | Total Technical Debt (2) | Maximum Nesting Level (1) | Number of Parameters (1) |
| Total Principal (4) | Access to Foreign Data (1) | Mean Message Chain Length (1) | Number of Private Attributes (1) |
| Anonymous Classes (3) | Access to Local Data (1) | Measure of Aggregation (1) | Number of Private Methods (1) |
| Cognitive Complexity (3) | Afferent Coupling (1) | Memory Heap (1) | Number of Protected Attributes (1) |
| Comment Lines Density (3) | Average Method Weight (1) | New Keyword Statements (1) | Number of Protected Methods (1) |
| Comparison (3) | Average Method Weight of Not Accessor or Mutator Methods (1) | Number of Abstract Methods (1) | Number of Public Attributes (1) |
| Duplicated Blocks (3) | Called Foreign Not Accessor or Mutator Methods (1) | Number of Accessed Variables (1) | Number of Static Attributes (1) |
| Math Operations (3) | Called Local Not Accessor or Mutator Methods (1) | Number of Accessor Methods (1) | Number of Static Methods (1) |
| Max Nested Blocks (3) | Changing Classes (1) | Number of Anonymous Functions (1) | Outgoing Module Dependency (1) |
| Non-empty Lines of Code (3) | Changing Methods (1) | Number of Arrow Functions (1) | Physical Source Code Lines (1) |
| Number of Fields (3) | Cohesion Among Methods (1) | Number of Closed Issues (1) | Popularity – Number of Stars (1) |
| Number of Unique Words (3) | Concern Overload (1) | Number of Co-changed Files (1) | Release Size in Bytes (1) |
| Parenthesized Expressions (3) | Contributors (1) | Number of Constructor Methods (1) | Reliability Remediation Effort (1) |
| SQALE Debt Ratio (3) | Coupling Between Methods (1) | Number of Default Attributes (1) | Reverse Days to the Latest Release (1) |
| String Literals (3) | Coupling Dispersion (1) | Number of Default Methods (1) | Scattered Functionality (1) |
| Total Number of Loops (3) | Coupling Intensity (1) | Number of Directories (1) | Security Remediation Effort (1) |
| Total Number of Modifiers (3) | Cross-module Co-changes (1) | Number of Files (1) | Source Code Coverage Percent (1) |
| Total Number of Numbers Used (3) | Cyclomatic Complexity Density (1) | Number of Final Methods (1) | Tight Class Cohesion (1) |
| Total Number of Returns (3) | Data Access Metric (1) | Number of Final and Non-Static Attributes (1) | Total Incoming Module Dependencies (1) |
| Total Number of Variables (3) | Efferent Coupling (1) | Number of Final and Non-Static Methods (1) | Total Outgoing Module Dependencies (1) |
| Try/Catches (3) | External Module Dependencies (1) | Number of Final and Static Attributes (1) | Unstable Dependency (1) |
| Uncovered Lines (3) | Fanout (1) | Number of Final and Static Methods (1) | Version of ECMAScript Applied (1) |
| Usage of Each Field (3) | Foreign Data Provider (1) | Number of Implemented Interfaces (1) | Weight of Class (1) |
| Usage of Each Variable (3) | Frequency of Updates (1) | Number of Inherited Methods (1) | Weighted Method Count of Not Accessor or Mutator Methods (1) |

| | |
|---|---|
| Comments per Commit (2) | |

Table X provides a comprehensive overview of the aggregate number of metrics employed in TD evolution forecasting research, along with the frequency of their utilization. The findings reveal that 13 metrics are predominantly utilized to address TD evolution forecasting (utilized five or more times in the conducted studies). These metrics include code smells, bugs, complexity, lines of code, vulnerabilities, comment lines, depth of inheritance tree, coupling between objects, lack of cohesion in methods, weight method count per class, response for a class, number of classes, and number of methods. Moreover, among these 13 metrics, code complexity and lines of code stand out as the most frequently employed. Additionally, we explained the top 15 TD indicator metrics that have been used in the literature, providing their definitions and explaining their relationships with forecasting, in Table XI.

Table XI: Top Metrics and Their Relationship to TD Forecasting

| **Metric** | **Definition** | **Relationship to TD Forecasting** |
|---|---|---|
| Complexity | The degree of complication within the code. | High complexity can indicate intricate code that may be harder to maintain, potentially leading to future TD. |
| Line of Code | The total number of lines in the source code. | A larger codebase may result in increased maintenance efforts, contributing to TD. |
| Depth of Inheritance Tree | The length of the inheritance path from a class to its root class. | Excessive inheritance depth might lead to complex hierarchies, impacting code maintainability and contributing to TD. |
| Bugs | Defects or errors in the code that need to be fixed. | The presence of bugs suggests existing issues that may contribute to TD. |
| Code Smells | Indications of poor coding practices that may lead to issues. | Indicative of poor code quality, addressing code smells can prevent the accumulation of TD. |
| Coupling Between Objects | The degree of dependency between classes or modules. | High coupling may lead to dependencies, making the codebase more susceptible to changes and potential TD. |
| Vulnerabilities | Weaknesses in the code that can be exploited for security breaches. | Security vulnerabilities can introduce TD in the form of potential future issues or breaches. |

Continued on next page

Table XI – continued from previous page

| Metric | Definition | Relationship to TD Forecasting |
|---|---|---|
| Lack of Cohesion in Methods | The degree to which methods within a class are unrelated. | Low cohesion may result in scattered and harder-to-maintain code, contributing to TD. |
| Number of Methods | The total count of methods within a class or module. | A higher number of methods might indicate a more complex class, potentially leading to higher maintenance efforts. |
| Response for a Class | Measures the responsiveness of a class to external requests. | Indicates how responsive a class is; low responsiveness may lead to future maintenance challenges. |
| Weight Method Count per Class | The total weight of methods within a class. | High method count in a class may contribute to increased complexity and maintenance efforts. |
| Comment Lines | The number of lines containing comments in the code. | Insufficient comments may make the code harder to understand and maintain, contributing to TD. |
| Number of Classes | The total count of classes in the codebase. | A higher number of classes may lead to increased complexity and potential TD. |
| Number of Static Invocations | The number of times static methods are called. | High static invocations may indicate a reliance on global state, potentially contributing to TD. |
| SQALE Index | Software Quality Assessment based on various factors. | SQALE Index provides an overall assessment of software quality; a lower index may indicate potential TD. |

*Answer to $RQ_4$:*
*It is evident that 13 metrics are prominently utilized for forecasting and predicting TD evolution, with code complexity and lines of code emerging as the most dominant metrics.*

### 4.6. *$RQ_5$: Which evaluation metrics are predominantly used for measuring the performance of TD forecasting and prediction models*

Evaluation metrics play a vital role in enabling developers and researchers to comprehensively assess the performance and practical applicability of developed models to real-world problems. In light of this, we meticulously identified nine distinct evaluation metrics employed thus far in the forecasting of TD evolution. These metrics included the Root Mean Square Error (RMSE), Mean Absolute Percentage Error (MAPE), Mean Absolute Error (MAE), Accuracy, F-Measure, Precision, Recall, Area Under the Curve (AUC), and Receiver Operating Characteristic (ROC).

The proportional utilization of these metrics is illustrated in Figure V. Furthermore, Figure V underscores the prominence of RMSE, MAPE, MAE, and F-measure in the domain of TD evolution forecasting models. This observation highlights the significant interest among researchers in quantifying the discrepancies between predicted and actual values as well as in gauging the percentage of these disparities.

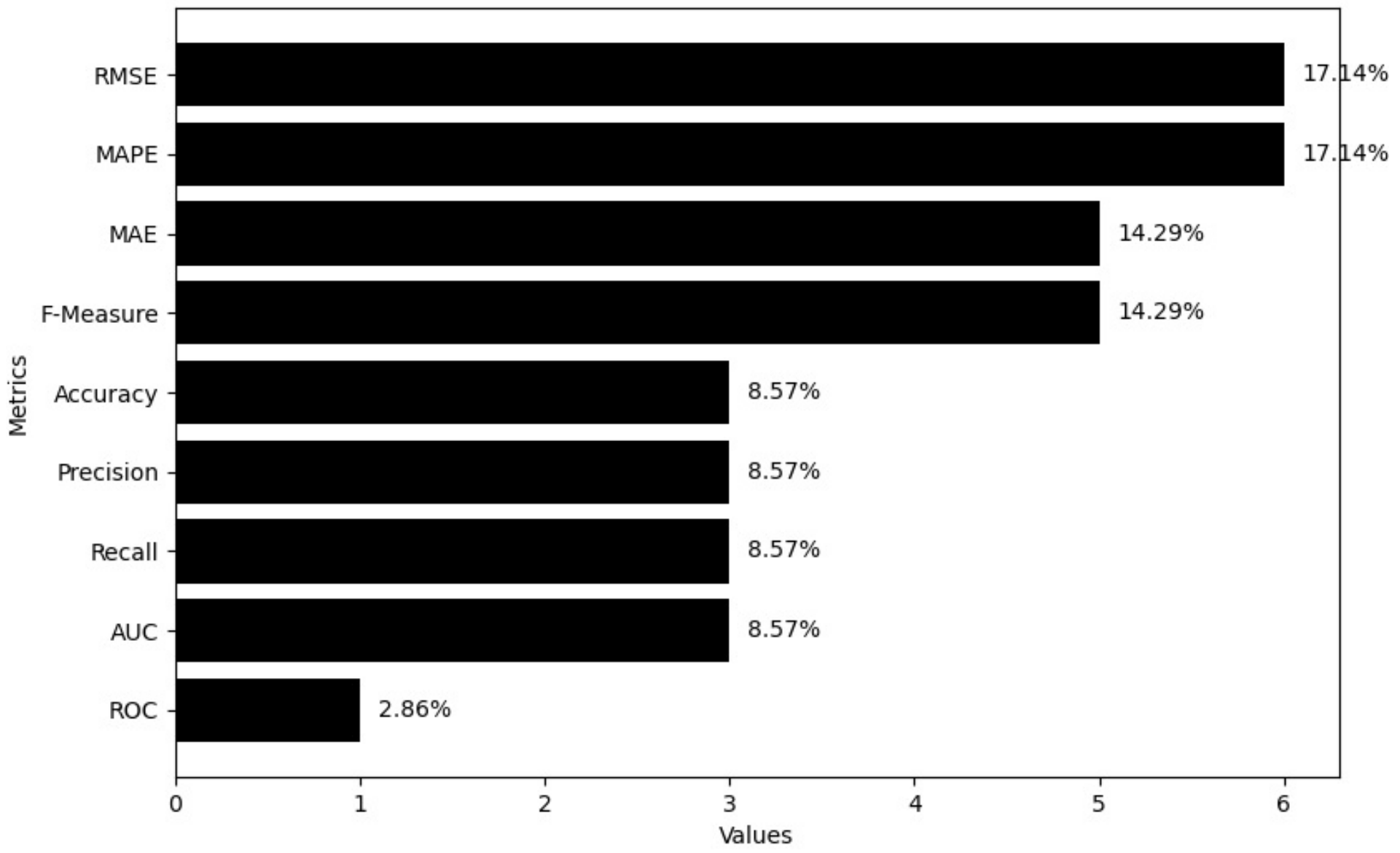

Figure V: Percentage of evaluation metrics used in forecasting of TD evolution

However, it's crucial to acknowledge that while Figure V provides insight into the individual prevalence of each metric across studies, it doesn't explicitly convey whether researchers used these metrics independently or in combination within their work. Each of these prominent identified metrics has individual limitations, including assumptions of independence, sensitivity to outliers, and a lack of interpretability. Therefore, we emphasize the importance for researchers not to rely exclusively on a single evaluation metric when assessing model performance. It is recommended to incorporate additional measures, such as the MSLE, which calculates the average logarithmic difference between predicted and actual values, and weighted metrics, which consider the importance or severity of different types of TD. By combining these evaluation metrics, researchers can gain a better understanding of the performance of the developed model.

*Answer to $RQ_5$:*
*Diverse set of nine evaluation metrics was employed in the context of TD evolution forecasting. Notably, RMSE, MAPE, MAE, and F-measure emerged as the most extensively utilized metrics in this domain.*

## 5. Discussion

In this section, we present the insights derived from our SLR on TD forecasting. Despite being a relatively young field compared to TD prioritization or management, the past five years have seen significant contributions, with a growing interest among researchers.

Machine learning and deep learning methods play a crucial role in identifying patterns in datasets, particularly when forecasting TD evolution. Our systematic literature review identified 23 distinct approaches, with 10 demonstrating notable success. According to the findings from the literature, Random Forests (RF) and Temporal Convolutional Networks (TCN) have emerged as particularly effective methods, possibly owing to their specific characteristics, including high accuracy, robustness, feature importance, versatility, and scalability (Ahsan et al., 2021). This result suggests a compelling avenue for researchers and practitioners. Incorporating these methods into forecasting models can enhance predictive capabilities, offering valuable insights into the complex dynamics of TD evolution. By harnessing the strengths of machine learning, software engineering professionals stand to benefit from more accurate, adaptable, and scalable solutions. This, in turn, aids in the proactive management and mitigation of TD challenges within evolving software systems.

Moreover, our exploration of the existing literature, in alignment with prior studies such as Li et al. (Li et al., 2015) and Lenarduzzi et al. (Lenarduzzi et al., 2021), highlights a concentration on two predominant types of TD: code debt and architectural debt. This focused attention on specific debt types is indicative of their tangible nature and the ease with which they can be measured. Code debt, rooted in suboptimal coding practices, directly impacts code quality and maintainability. On the other hand, architectural debt, revolving around design decisions, exerts influence over scalability and system extensibility. However, this targeted emphasis on code and architectural debt might be viewed as a limitation in the existing research landscape. Notably, our findings reveal that the remaining 13 types of TD have received limited attention. This glaring gap in research suggests a significant challenge for software engineers in effectively managing and mitigating associated risks. The lack of exploration into these less-addressed types of TD poses potential threats to software quality and development productivity, emphasizing the need for future research endeavors to comprehensively address the entire spectrum of TD types. Such endeavors would contribute not only to a more holistic understanding of TD but also to the development of strategies that encompass the broader landscape of challenges faced by software engineering professionals.

Similarly, crucial to training and validating forecasting approaches is the choice of datasets. We identified 66 software benchmarks in TD evolution research, with 10 benchmarks prominently featured, such as Commons-io, Commons-codec, Zookeeper, Jackson-dataformat-xml, Mina-sshd, Common-imaging, Jfreechart, Httpcomponents-client, Jackson-core, and Square OkHttp. These benchmarks, predominantly in Java, offer diverse software contexts for comprehensive TD evolution analysis. The implication of this result is that training and validating TD forecasting models on a diverse set of benchmarks are better equipped to provide actionable insights and predictions relevant to the challenges faced by software engineers. This adaptability can enhance the applicability of these forecasting approaches in real-world settings, offering valuable support in the ongoing efforts to manage and mitigate TD effectively.

In the same way, an array of metrics shapes the performance of TD evolution forecasting models. Our review identified 153 metrics, with 13 being predominantly used, including code smells, bugs, complexity, line of code, vulnerabilities, comment lines, depth of inheritance tree, coupling between objects, lack of cohesion in methods, weight method count per class, response for a class, number of classes, and number of methods. Among these metrics, code complexity and line of code are particularly prevalent. Metrics such as code complexity, line of code, code smells, bugs, and vulnerabilities play vital roles in assessing software quality, maintainability, and security. This suggests to software engineers that the selection of appropriate metrics is paramount in developing effective TD evolution forecasting models. Software engineering practitioners should understand the significance of specific metrics like code complexity and line of code; these metrics might serve as key inputs in forecasting models. The emphasis on these metrics in the literature and our study suggests that software engineers should prioritize their consideration when developing and implementing strategies to manage TD.

Finally, this study shows that out of all the evaluation metrics that have been employed in the existing TD forecasting literature, the evaluation metrics like RMSE, MAPE, MAE, and F-measure provide comprehensive assessments of forecasting performance, as they are the most used evaluation metrics among the researchers. Researchers and software engineers should note that the choice of evaluation metrics is critical in accurately gauging the performance of TD forecasting models. RMSE (Root Mean Square Error) is particularly valuable as it measures the average magnitude of the forecasting errors, providing an overall indication of the model's accuracy. MAPE (Mean Absolute Percentage Error) evaluates the accuracy of predictions in percentage terms, offering insights into the scale of errors relative to the actual values. MAE (Mean Absolute Error) provides a straightforward measure of forecasting accuracy by calculating the average absolute errors between predicted and actual values. F-measure, a metric that combines precision and recall, is crucial for assessing the model's ability to balance between false positives and false negatives.

The prevalence of these metrics in the literature highlights their importance in capturing different facets of forecasting performance. Consequently, researchers and software engineers are encouraged to consider a combination of these metrics to obtain a well-rounded evaluation of TD forecasting models. This approach ensures a more understanding of the strengths and limitations of the models, supporting informed decision-making in the development and deployment of TD forecasting solutions in real-world software engineering scenarios.

*Future research*

Through an analysis of existing studies, we identified the following areas for future research.

**Assess the generalizability and transferability of Random Forest and Temporal Convolutional Network models across different software domains and contexts:** It is crucial to thoroughly investigate whether Random Forest and Temporal Convolutional Network models consistently outperform other approaches across various software solutions, coding languages, and development methodologies. The reason is that in certain studies, these models have demonstrated remarkable performance and have been considered top performers. Additionally, exploring transfer learning strategies becomes significant to leverage pre-trained models from one software domain and apply them effectively in related areas where there is limited labeled data available. Such research could provide valuable insights into the adaptability and versatility of temporal convolutional networks and random forest models, enabling their potential utilization beyond the scope of specific software systems.

**Predicting the Consequences of Non-code and Non-architecture TD evolution on Software Reliability:** This research aims to investigate the impact of non-code and non-architecture debts on software reliability, exploring their effects on overall software system quality assurance. Such research could provide novel forecasting techniques to predict the evolution of these debts and provides insights for effective management and mitigation strategies to enhance software reliability.

**Expanding the coverage of software benchmarks for evaluating the forecasting of TD evolution models:** The findings of this research suggest the need to expand the coverage of benchmarks used in TD evolution forecasting research beyond Java and JavaScript. It is important to investigate benchmarks written in other programming languages to gain a more comprehensive understanding of the applicability and limitations of the existing models across diverse software ecosystems.

**Investigating the relationship between the identified metrics and their impact on TD evolution:** Further research can explore the specific cause-and-effect relationships between the 13 identified metrics (such as code smells, bugs, complexity, etc.) and the subsequent evolution of TD. This can help to establish a deeper understanding of how these metrics contribute to TD and provide insights into effective mitigation strategies.

**Developing forecasting and predictive models and algorithms for TD evolution:** Building upon the identified 13 metrics, future research can aim to develop robust predictive and forecasting models and algorithms for TD evolution. These models can utilize machine learning, data mining, or statistical techniques to forecast TD trends and help software development teams prioritize and allocate resources for effective TD management.

## 6. Threats to validity

### *6.1. Internal Validity*

To ensure the internal validity of our study, we conducted a thorough validation of the entire selection process, placing particular emphasis on the data extraction protocol. All authors actively participated in this validation process to guarantee that the schema aligned appropriately with the defined research objective and corresponding research questions. Rigorous verification of the extracted data from each paper was carried out before answering the research questions, minimizing the potential for errors in the data extraction process and ensuring the internal consistency of our study.

### *6.2. External Validity*

In addressing threats to external validity, we scrutinized the entire SLR process to minimize potential theoretical concerns. The selection processes were executed by a minimum of two researchers, and any disagreements were extensively discussed to enhance the reliability of the study. Refinements to the inclusion and exclusion criteria were made to improve objectivity. While the search string might not have covered all relevant terms, we conducted pilot tests and used the Connected Paper AI tool to refine the search string and ensure comprehensive coverage within the selected time periods. This approach enhances the generalizability of our findings beyond the sampled studies, strengthening the external validity of our study.

### *6.3. Construct Validity*

To ensure construct validity, we recognized the potential correlation between the failure to identify and include primary studies and the loss of relevant evidence. Mitigating this threat, we adopted a comprehensive approach by considering multiple data sources, thereby minimizing the risk of omitting pertinent studies. Specifically targeting reputable scholarly sources and digital libraries in computer science and software engineering enhanced the relevance of the studies included in our review. Diligent search strategies were implemented to retrieve a significant number of articles closely aligned with the study's objective, contributing to the construct validity of our findings.

## 7. Conclusion

This study provides a comprehensive overview of the current state of research on TD forecasting. These findings emphasize the growing interest and activity in this field, as researchers increasingly acknowledge the significance of comprehending and handling TD in software development projects. This study identifies various approaches used for TD forecasting, with Random Forests (RF) and a Temporal Convolutional Network (TCN), which are particularly effective methods owing to their ability to capture the complexities and temporal dynamics of TD evolution.

This study sheds light on the types of debt and datasets commonly examined in TD evolution forecasting. Code debt and architectural debt receive the most attention, likely because they are more tangible and easier to measure than other types of TD. Moreover, the study reveals a focus on Java benchmarks, reflecting the popularity and prevalence of the Java programming language in the software industry.

Furthermore, the selection of metrics and evaluation criteria plays a vital role in developing and assessing TD evolution forecasting models. Metrics such as code complexity, line of code, code smells, bugs, vulnerabilities, and various others have been extensively used to capture different aspects of software quality and maintainability. Evaluation metrics such as RMSE, MAPE, MAE, and F-measure are commonly employed to assess the accuracy and performance of TD evolution forecasting approaches.

While significant progress has been made in TD evolution forecasting research, there are several avenues for future exploration. This study suggests investigating the generalizability and transferability of RF and TCN models across software domains, coding languages, and development methodologies to determine their consistent performance. It also recommends expanding the coverage of benchmarks beyond Java and JavaScript to evaluate the TD evolution models in diverse software ecosystems. Additionally, future research should focus on exploring the cause-and-effect relationships between the identified metrics and TD evolution, as well as developing robust predictive models and algorithms for TD evolution using machine learning, data mining, and statistical techniques.

Overall, the research presented in this study contributes to advancing the understanding of TD forecasting, and provides a foundation for further research and practical applications in managing TD in software development. By addressing the identified research gaps and pursuing suggested future directions, researchers and practitioners can make informed decisions, allocate resources effectively, and mitigate the potential risks associated with TD in software systems.

**Author Contributions:** Conceptualization, A.A,; methodology, A.A, A.Y, and A.T; data curation, A.A, A.Y and A.T; original draft preparation, A.A; writing-review and editing, A.A, A.Y, and A.T visualization, A.A; All authors have read and agreed to the published version of the manuscript.

**Funding:** This work did not receive fund from any organization

**Conflicts of Interest:** The authors declare no conflict of interest.

## Primary studies

Ardimento, P., Aversano, L., Bernardi, M. L., Cimitile, M., and Iammarino, M. (2022). Using deep temporal convolutional networks to just-in-time forecast technical debt principal. *Journal of Systems and Software*, 194:111481.

Aversano, L., Bernardi, M. L., Cimitile, M., and Iammarino, M. (2021). Technical debt predictive model through temporal convolutional network. In *2021 International Joint Conference on Neural Networks (IJCNN)*, pages 1–8. IEEE.

Aversano, L., Bernardi, M. L., Cimitile, M., Iammarino, M., and Montano, D. (2023). Forecasting technical debt evolution in software systems: an empirical study. *Frontiers of Computer Science*, 17(3):173210.

Fontana, F. A., Avgeriou, P., Pigazzini, I., and Roveda, R. (2019). A study on architectural smells prediction. In *2019 45th Euromicro Conference on Software Engineering and Advanced Applications (SEAA)*, pages 333–337. IEEE.

Garcia, J., Kouroshfar, E., Ghorbani, N., and Malek, S. (2021). Forecasting architectural decay from evolutionary history. *IEEE Transactions on Software Engineering*, 48(7):2439–2454.

Lerina, A., Bernardi, M. L., Cimitile, M., and Iammarino, M. (2022). Technical debt forecasting from source code using temporal convolutional networks. In *Product-Focused Software Process Improvement: 23rd International Conference, PROFES 2022, Jyväskylä, Finland, November 21–23, 2022, Proceedings*, pages 581–591. Springer.

Mathioudaki, M., Tsoukalas, D., Siavvas, M., and Kehagias, D. (2021). Technical debt forecasting based on deep learning techniques. In *Computational Science and Its Applications–ICCSA 2021: 21st International Conference, Cagliari, Italy, September 13–16, 2021, Proceedings, Part VII 21*, pages 306–322. Springer.

Mathioudaki, M., Tsoukalas, D., Siavvas, M., and Kehagias, D. (2022). Comparing univariate and multivariate time series models for technical debt forecasting. In *Computational Science and Its Applications–ICCSA 2022 Workshops: Malaga, Spain, July 4–7, 2022, Proceedings, Part IV*, pages 62–78. Springer.

Mhawish, M. Y. and Gupta, M. (2020). Predicting code smells and analysis of predictions: Using machine learning techniques and software metrics. *Journal of Computer Science and Technology*, 35:1428–1445.

Rantala, L. and Mäntylä, M. (2020). Predicting technical debt from commit contents: reproduction and extension with automated feature selection. *Software Quality Journal*, 28:1551–1579.

Tsoukalas, D., Jankovic, M., Siavvas, M., Kehagias, D., Chatzigeorgiou, A., and Tzovaras, D. (2019). On the applicability of time series models for technical debt forecasting. In *15th China-Europe International Symposium on software engineering education*.

Tsoukalas, D., Kehagias, D., Siavvas, M., and Chatzigeorgiou, A. (2020). Technical debt forecasting: an empirical study on open-source repositories. *Journal of Systems and Software*, 170:110777.

Tsoukalas, D., Mathioudaki, M., Siavvas, M., Kehagias, D., and Chatzigeorgiou, A. (2021). A clustering approach towards cross-project technical debt forecasting. *SN Computer Science*, 2(1):22.

Zozas, I., Bibi, S., and Ampatzoglou, A. (2022). Forecasting the principal of code technical debt in javascript applications. *IEEE Transactions on Software Engineering*.

## References

Ahsan, M., Gomes, R., Chowdhury, M. M., and Nygard, K. E. (2021). Enhancing machine learning prediction in cybersecurity using dynamic feature selector. *Journal of Cybersecurity and Privacy*, 1(1):199–218.

Alfayez, R., Alwehaibi, W., Winn, R., Venson, E., and Boehm, B. (2020). A systematic literature review of technical debt prioritization. In *Proceedings of the 3rd International Conference on Technical Debt*, pages 1–10.

Allman, E. (2012). Managing technical debt: Shortcuts that save money and time today can cost you down the road. *Queue*, 10(3):10–17.

Alpaydin, E. (2010). Design and analysis of machine learning experiments.

Alves, N. S., Mendes, T. S., de Mendonça, M. G., Spínola, R. O., Shull, F., and Seaman, C. (2016). Identification and management of technical debt: A systematic mapping study. *Information and Software Technology*, 70:100–121.

Amanatidis, T., Mittas, N., Moschou, A., Chatzigeorgiou, A., Ampatzoglou, A., and Angelis, L. (2020). Evaluating the agreement among technical debt measurement tools: building an empirical benchmark of technical debt liabilities. *Empirical Software Engineering*, 25:4161–4204.

Ampatzoglou, A., Ampatzoglou, A., Chatzigeorgiou, A., and Avgeriou, P. (2015). The financial aspect of managing technical debt: A systematic literature review. *Information and Software Technology*, 64:52–73.

Avgeriou, P., Kruchten, P., Nord, R. L., Ozkaya, I., and Seaman, C. (2015). Reducing friction in software development. *Ieee software*, 33(1):66–73.

Behutiye, W. N., Rodríguez, P., Oivo, M., and Tosun, A. (2017). Analyzing the concept of technical debt in the context of agile software development: A systematic literature review. *Information and Software Technology*, 82:139–158.

Belle, A. B. (2019). Estimation and prediction of technical debt: a proposal. *arXiv preprint arXiv:1904.01001*.

Besker, T., Martini, A., and Bosch, J. (2018). Managing architectural technical debt: A unified model and systematic literature review. *Journal of Systems and Software*, 135:1–16.

Bogner, J., Verdecchia, R., and Gerostathopoulos, I. (2021). Characterizing technical debt and antipatterns in ai-based systems: A systematic mapping study. In *2021 IEEE/ACM International Conference on Technical Debt (TechDebt)*, pages 64–73. IEEE.

Bontempi, G., Ben Taieb, S., and Le Borgne, Y.-A. (2013). Machine learning strategies for time series forecasting. *Business Intelligence: Second European Summer School, eBISS 2012, Brussels, Belgium, July 15-21, 2012, Tutorial Lectures 2*, pages 62–77.

Brown, N., Cai, Y., Guo, Y., Kazman, R., Kim, M., Kruchten, P., Lim, E., MacCormack, A., Nord, R., Ozkaya, I., et al. (2010). Managing technical debt in software-reliant systems. In *Proceedings of the FSE/SDP workshop on Future of software engineering research*, pages 47–52.

Christy, O. N., Umar, Y. H., and Agbailu, A. O. (2022). A comparative study of autoregressive integrated moving average and artificial neural networks models. *Studies*, 5(3):54–74.

Cunningham, W. (1992). The wycash portfolio management system. *ACM SIGPLAN OOPS Messenger*, 4(2):29–30.

Das, J. (2012). *Statistics for business decisions*. Academic Publishers.

Digkas, G., Lungu, M., Chatzigeorgiou, A., and Avgeriou, P. (2017). The evolution of technical debt in the apache ecosystem. In *Software Architecture: 11th European Conference, ECSA 2017, Canterbury, UK, September 11-15, 2017, Proceedings 11*, pages 51–66. Springer.

Dybå, T. and Dingsøyr, T. (2008). Empirical studies of agile software development: A systematic review. *Information and software technology*, 50(9-10):833–859.

Ernst, N., Kazman, R., and Delange, J. (2021). *Technical Debt in Practice: How to Find It and Fix It*. MIT Press.

Fernández-Sánchez, C., Garbajosa, J., Yagüe, A., and Perez, J. (2017). Identification and analysis of the elements required to manage technical debt by means of a systematic mapping study. *Journal of Systems and Software*, 124:22–38.

Friedman, J., Hastie, T., and Tibshirani, R. (2001). The elements of statistical learning. vol. 1 springer series in statistics. *New York*.

Khomyakov, I., Makhmutov, Z., Mirgalimova, R., and Sillitti, A. (2019). Automated measurement of technical debt: A systematic literature review. In *ICEIS 2019-Proceedings of the 21st International Conference on Enterprise Information Systems*, pages 95–106.

Kitchenham, B. and Brereton, P. (2013). A systematic review of systematic review process research in software engineering. *Information and software technology*, 55(12):2049–2075.

Kitchenham, B. and Charters, S. (2007). Guidelines for performing systematic literature reviews in software engineering. 2.

Kitchenham, B. A. (2012). Systematic review in software engineering: where we are and where we should be going. In *Proceedings of the 2nd international workshop on Evidential assessment of software technologies*, pages 1–2.

Kitchenham, B. A., Budgen, D., and Brereton, P. (2015). *Evidence-based software engineering and systematic reviews*, volume 4. CRC press.

Lenarduzzi, V., Besker, T., Taibi, D., Martini, A., and Fontana, F. A. (2021). A systematic literature review on technical debt prioritization: Strategies, processes, factors, and tools. *Journal of Systems and Software*, 171:110827.

Li, Z., Avgeriou, P., and Liang, P. (2015). A systematic mapping study on technical debt and its management. *Journal of Systems and Software*, 101:193–220.

Makridakis, S., Spiliotis, E., and Assimakopoulos, V. (2018). Statistical and machine learning forecasting methods: Concerns and ways forward. *PloS one*, 13(3):e0194889.

McConnell, S. (2008). Managing technical debt. *Construx Software Builders, Inc*, pages 1–14.

Palit, A. K. and Popovic, D. (2006). *Computational intelligence in time series forecasting: theory and engineering applications*. Springer Science & Business Media.

Ribeiro, L. F., de Freitas Farias, M. A., Mendonça, M. G., and Spínola, R. O. (2016). Decision criteria for the payment of technical debt in software projects: A systematic mapping study. *ICEIS (1)*, pages 572–579.

Rios, N., de Mendonça Neto, M. G., and Spínola, R. O. (2018). A tertiary study on technical debt: Types, management strategies, research trends, and base information for practitioners. *Information and Software Technology*, 102:117–145.

Said, K. S., Nie, L., Ajibode, A. A., and Zhou, X. (2020). Gui testing for mobile applications: objectives, approaches and challenges. In *Proceedings of the 12th Asia-Pacific Symposium on Internetware*, pages 51–60.

Seaman, C. and Guo, Y. (2011). Measuring and monitoring technical debt. In *Advances in Computers*, volume 82, pages 25–46. Elsevier.

Sierra, G., Shihab, E., and Kamei, Y. (2019). A survey of self-admitted technical debt. *Journal of Systems and Software*, 152:70–82.

Skourletopoulos, G., Mavromoustakis, C. X., Bahsoon, R., Mastorakis, G., and Pallis, E. (2014). Predicting and quantifying the technical debt in cloud software engineering. In *2014 IEEE 19th international workshop on computer aided modeling and design of communication links and networks (CAMAD)*, pages 36–40. IEEE.

Tom, E., Aurum, A., and Vidgen, R. (2013). An exploration of technical debt. *Journal of Systems and Software*, 86(6):1498–1516.

Tsoukalas, D., Siavvas, M., Jankovic, M., Kehagias, D., Chatzigeorgiou, A., and Tzovaras, D. (2018). Methods and tools for td estimation and forecasting: A state-of-the-art survey. In *2018 International Conference on intelligent systems (IS)*, pages 698–705. IEEE.

Wohlin, C. (2014). Guidelines for snowballing in systematic literature studies and a replication in software engineering. In *International Conference on Evaluation & Assessment in Software Engineering*.

Wohlin, C., Runeson, P., Höst, M., Ohlsson, M. C., Regnell, B., and Wesslén, A. (2012). *Experimentation in software engineering*. Springer Science & Business Media.